\documentclass[structabstract]{aa}
\usepackage{graphicx}
\usepackage{txfonts}
\begin{document}

\title{A fresh insight into the evolutionary status and third body hypothesis of the eclipsing binaries \object{AD And}, \object{AL Cam}, and \object{V338 Her}}

\author{A. Liakos\inst{1} \and P. Niarchos\inst{1} \and E. Budding\inst{2,3}}

\institute{University of Athens, Department of Astrophysics, Astronomy and Mechanics, GR 157 84, Zografos, Athens, Hellas\\
\email{alliakos@phys.uoa.gr, pniarcho@phys.uoa.gr}
\and
Carter Observatory \& School of Chemical and Physical Sciences, Victoria University, Wellington, New Zealand
\and
Department of Physics and Astronomy, University of Canterbury, Christchurch, New Zealand\\
\email{budding@xtra.co.nz}}

\date{Received September XX, XXXX; accepted March XX, XXXX}


\abstract
{}
{We aim to derive the absolute parameters of the components of AD~And, AL~Cam, and V338~Her, interpret their orbital period changes and discuss their evolutionary status.}
{New and complete multi-filter light curves of the eclipsing binaries AD~And, AL~Cam, and V338~Her were obtained and analysed with modern methods. Using all reliably observed times of minimum light, we examined orbital period irregularities using the least squares method. In addition, we acquired new spectroscopic observations during the secondary eclipses for AL~Cam and V338~Her.}
{For AL~Cam and V338~Her, we derive reliable spectral types for their primary stars. Statistical checks of orbital period analysis for all systems are very reassuring in the cases of V338~Her and AD And, although less so for AL~Cam. The LIght-Time Effect (LITE) results are checked by inclusion of a third light option in the photometric analyses. Light curve solutions provide the means to calculate the absolute parameters of the components of the systems and reliably estimate their present evolutionary status.}
{AL~Cam and V338~Her are confirmed as classical Algols of relatively low mass in similar configurations. Unlike AL~Cam, however, V338~Her is still transferring matter between its components, raising interest in the determinability of the evolutionary histories of Algols. AD~And is found to be a detached system in which both close stars are of age $\sim10^9$~yr and is probably a `non-classical' young triple, at an interesting stage of its dynamical evolution.  }


\titlerunning{Analysis of the eclipsing binaries AD~And, AL~Cam, and V338~Her}
\authorrunning{A. Liakos, P. Niarchos and E. Budding}
\maketitle
%

\section{Introduction}

One main purpose of the present investigation is to derive new physical elements for AD~And, AL~Cam, and V338~Her based on new photometric and spectroscopic data and, using up-to-date analysis tools, examine their orbital period data. The particular systems studied were chosen for two main reasons: (1) they have known irregularities in their `observed $-$ calculated' times of minima (hereafter O$-$C), and (2) their literature light curves are either incomplete, or non-multi-filtered. New light curves were sought: (a) to search photometrically for additional companions of the eclipsing binary (hereafter EB), and (b) to attempt to detect the possible pulsational behaviour of its components. The O$-$C diagram analysis leads us to consider the physical mechanisms affecting an EB's behaviour, (cf. Budding \& Demircan \cite{BD07}, Ch.~8). These might be related to either the presence of a third body, quadrupole moment variations, mass transfer between the components, or mass loss from the system. At the same time, parameters derived from the light curve (hereafter LC) analysis allow a useful specification of the evolutionary status of EB stars (e.g. in relation to semi-detached, detached, or contact configurations).

Brancewicz \& Dworak (\cite{BD80}) produced absolute elements for the binary components studied in this paper, including photometric parallaxes. However, our new LCs, spectroscopic information, and data processing provide results of higher accuracy. Two of the systems, AL~Cam and V338~Her, were noted as candidates containing $\delta$~Sct type components in the catalogue of Soydugan et al. (\cite{SO06}), while AD~And and AL~Cam were classified by Hoffman et al. (\cite{HO06}) as possible triples.

\paragraph{AD And:}
The light variability of this $\beta$~Lyr-type{\footnote{This historically used term strictly corresponds to the LC only, and physical comparisons should not be made with the complex and massive $\beta$-Lyr itself.}} system was detected by Lacchini (\cite{LA27}). The first (visual) photometric study was that of Taylor \& Alexander (\cite{TA40}), while Rucinski (\cite{RU66}) published the first photoelectric LC. Cannon (\cite{CA34}) classified AD~And as an F-type object, although the later MK classification of Hill et al. (\cite{HI75}) gave the type in the range from B8 to A0. A period analysis was published by Frieboes-Conde \& Herczeg (\cite{FH73}), that gave a rather unreliable mass for the inferred third body. Str\"{o}mgren indices were given by Hilditch \& Hill (\cite{HH75}). Giuricin \& Mardirossian (\cite{GM81}) published revised photometric elements, and later AD~And was included by Hegedus (\cite{HE88}) in a list of systems with apparently displaced secondary minima, indicating an eccentric orbit. Samolyk (\cite{SA97}), however, carried out a new period study, based on the most modern times of minima, and found no period change. On the other hand, Liao \& Qian (\cite{LQ09}), following further O$-$C analysis, reported a LITE variation with a 14.38~yr period.

\begin{table*}
\centering
\caption{The photometric observations log and the reference (mean out-of-eclipse) magnitude data of the systems.}
\label{tab1}
\begin{tabular}{l cc cc cc cccc}
\hline
\hline
System&\multicolumn{2}{c}{Ref. mag}&  Nights&  Time span    &         \multicolumn{2}{c}{Filters used:}  &          Comparison   &  $V$ &           Check       & $V$  \\
                &   $B$   &   $V$  &  spent &     (MM/YY)   &\multicolumn{2}{c}{Points/Stan. Dev. (mmag)}&          star         & (mag)&           star        &(mag) \\
\hline
AD And          &  11.20  &  11.00 &   5    &   8/08-9/08   &       $B$:358/4.8 &$V$:344/5.0             &\object{GSC 3641-0045} &10.8  &\object{GSC 3641-0037} & 11.3 \\
                &         &        &        &               &       $R$:349/4.9 &$I$:349/4.2             &                       &      &                       &      \\
AL Cam          &  10.52  &  10.21 &   7    &   2/08-4/08   &       $B$:1304/4.2&$V$:1297/4.9            &\object{GSC 4556-0163} & 9.8  &\object{GSC 4556-0871} & 9.4  \\
V338 Her        &  10.42  &  10.07 &  13    &   5/08-8/08   &       $B$:742/4.6 &$V$:756/4.5             &\object{GSC 3101-0995} &10.7  &\object{GSC 3101-1186} & 12.2 \\
                &         &        &        &               &       $R$:743/4.9 &$I$:759/4.5             &                       &      &                       &      \\
\hline
\end{tabular}
\end{table*}

\paragraph{AL Cam:}
This Algol-type EB was discovered as a variable by Strohmeier (\cite{ST58}), who published the first photographic LC. Quester \& Braune (\cite{QB65}) produced an updated light ephemeris and Hilditch \& Hill (\cite{HH75}) presented Str\"{o}mgren indices of the system. The MK classification by Hill et al. (\cite{HI75}) yielded a spectral type in the range A4V-A7V. Srivastava (\cite{SR91}) carried out an O$-$C study of AL Cam and concluded that period changes on varying timescales are present, while Samolyk (\cite{SA96}) published an improved ephemeris.

\paragraph{V338 Her:}
This system was discovered as a variable, that has an Algol-type LC, by Hoffmeister (\cite{HO49}). Tsesevitch (\cite{TS51}) published light elements based on his photographic and visual observations. The first photoelectric LC and period study was given by Vete\v{s}nik (\cite{VE68}), while a set of geometric elements for the components were produced by Walter (\cite{WA69}). Hill et al. (\cite{HI75}) published the spectral type as F0V-F2V. The distance of V338~Her was reported as about 377~pc (Perryman \cite{PE97}). The quoted spectral type appears to have settled at F2V (SIMBAD). Yang et al. (\cite{YA10}) analysed the LCs and found the secondary to be somewhat `undersized', i.e. not filling its Roche lobe (cf. Kopal \cite{KO59}). Additionally, they carried out an orbital period analysis, reporting a cyclic variation with timescale 29.07~yr.

\section{Observations and data reduction}
The photometric observations discussed herein were gathered at the Gerostathopoulion Observatory of the University of Athens in 2008 (for details see Table \ref{tab1}), using the 0.4~m Cassegrain telescope equipped with an ST-8XMEI CCD camera and $BVRI$ (Bessell specification) photometric filters. Aperture photometry was applied to the raw data and differential magnitudes were obtained using the software \emph{MuniWin} v.1.1.23 (Hroch \cite{HR98}). Exposure times were arranged in order to search for any short-period pulsations. Further details of the comparison and check stars of each programme are given in Table \ref{tab1}.

In Table \ref{tab1}, we list reference $B$ and $V$ magnitudes for the mean out-of-eclipse light levels of the three binaries. In the case of AL~Cam, its main comparison star GSC~04556-00163 has SIMBAD values of $B$=10.30~mag and $V$=9.81~mag, so that our observed differential magnitudes lead directly to the values of Table \ref{tab3}. We note that $B-V$=0.31 is then consistent with the A8 type classification derived below. For the other two binaries, we obtained the $V$ reference values from Hilditch \& Hill (\cite{HH75}) and interpolated from the $b-y$ values given in that reference, checked also against the $J$ magnitude values given by SIMBAD for the comparison stars, using the linear gradient formula (cf. Budding \& Demircan \cite{BD07}, Ch.~3.6).

For V388~Her, the resulting $B$ and $V$ magnitudes agree with those given by SIMBAD. For AD~And, there is a disparity with the $V$ magnitude (11.2) listed by SIMBAD, but that appears to be an old photographic value and is inconsistent with the expected degree of reddening for that star (see below).

Spectroscopic observations of AL~Cam and V338~Her were obtained with the 1.3~m Ritchey-Cretien telescope at Skinakas Observatory, Crete Is. (Hellas), on 8 and 14 May 2009. A 2000$\times$800 ISA SITe CCD camera attached to a focal reducer with a 2400~lines/mm grating and slit of 80~$\mathrm{\mu}$m was used. This arrangement gave a nominal dispersion of 0.55~\AA/pixel and wavelength coverage of 4534-5622~\AA. Data were reduced with the \emph{AIP4WIN} software (Berry \& Burnell \cite{BB00}). The frames were bias-subtracted, a flat-field correction was applied, and the sky background was removed. The spectral region was selected so as to include H$_{\beta}$ and sufficient metallic lines. Before and after each on-target observation, an arc calibration exposure (NeHeAr) was recorded.

\section{Spectroscopic analysis}

A total of 19 spectroscopic standard stars, proposed by GEMINI Observatory{\footnote{http://www.gemini.edu/}}, ranging from A0 to G8 spectral types, were observed with the same instrumental set-up. Exposure times for the variables were 1800~s. All spectra were calibrated and normalized to enable direct comparisons. The spectra were then shifted, using H$_{\beta}$ as reference, to compensate for the relative Doppler shifts of each standard. The spectral region between 4800~\AA~and 5350~\AA, where H$_{\beta}$ and numerous metallic lines are strong was used for the spectral classification. The remaining spectral regions were ignored, because they generally lacked sufficient metallic lines with significant signal-to-noise ratios. The variables' spectra were subtracted from those of each standard, deriving sums of squared residuals in each case. These least squares sums guided us to the closest match between the spectra of variables and standards with a formal error of one subclass.
The spectra of the variables were also compared with synthetic spectra, following the same method, by using the \textit{SPECTRUM} software (Gray \& Corbally \cite{GC94}). The resolution of the synthetic spectra was chosen to be the same as that of the variables' spectra, while parameters such as microturbulence and macroturbulence were given typical values for this kind of stars. The effective temperatures and gravity parameters for the synthetic spectra were set to lie around the expected range of values of those of the variables.

\begin{figure*}[t]
\centering
\begin{tabular}{cl}
\includegraphics[width=15cm]{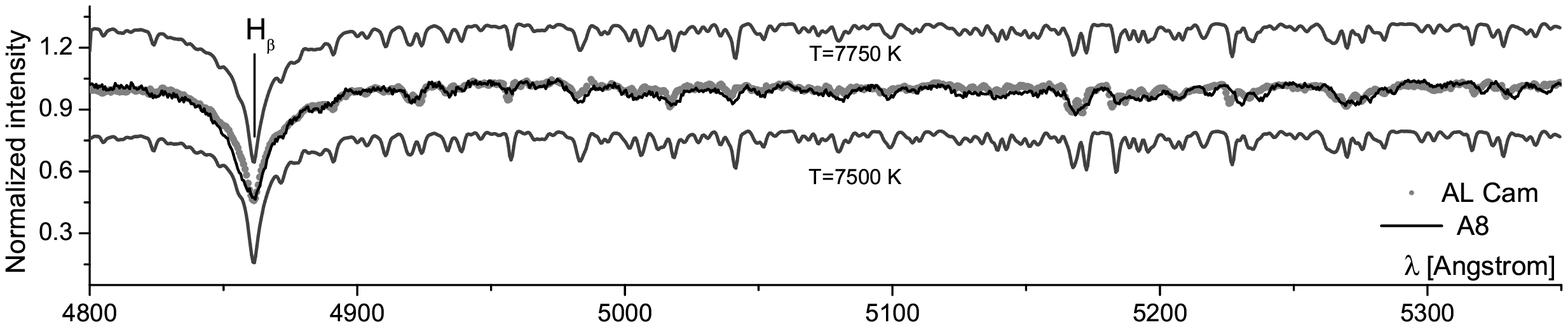}&(a)\\
\includegraphics[width=15cm]{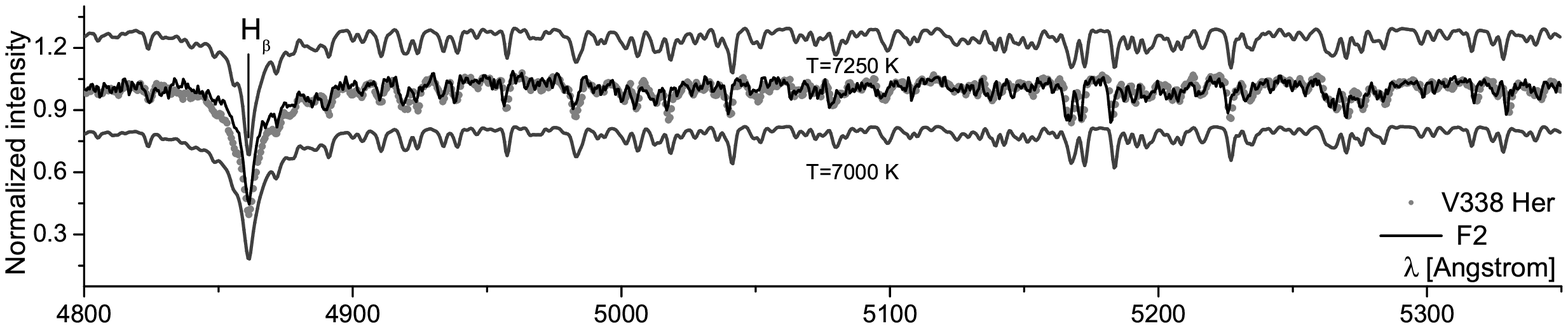}&(b)
\end{tabular}
\caption{(a) The comparison spectrum of AL~Cam (points) and the standard star (black solid line). Synthetic spectra (dark grey lines) of similar temperature and $\log g=4$ are also indicated. (b) The same for V338~Her.}
\label{fig1}
\end{figure*}

AL~Cam was observed at phase 0.524 in order to minimize the light contribution from the cooler component. Using the standard star method we found that the primary is an A8$\pm$1V type star, while in synthetic spectra comparison we found a closest match for $\log g=4$ and the temperature range 7500-7750~K. Fig.~\ref{fig1}a shows the best matching of AL~Cam which was found to be with HIP~21273.

V338~Her was observed at phase 0.497 (secondary minimum), again to allow us to focus on the primary component. The closest spectral match was achieved with HIP~82020, which is of spectral type F2$\pm$1V. The best comparison with synthetic spectra corresponds to temperatures of 7000-7250~K and $\log g=4$ (see Fig.~\ref{fig1}b), with the 7000~K model providing the best agreement.

For both cases, the standard spectra fit those of the variables more closely than the synthetic ones that were used as additional confirmation of the spectral classification. Moreover, the temperature formal errors in the method based on the standards cover the entire temperature range of the synthetic spectra. We therefore decided to use the temperature calibration of the spectral types for further modelling.

\section{Light curve analysis}

\begin{figure}
\centering
\begin{tabular}{cl}
\includegraphics[width=8cm]{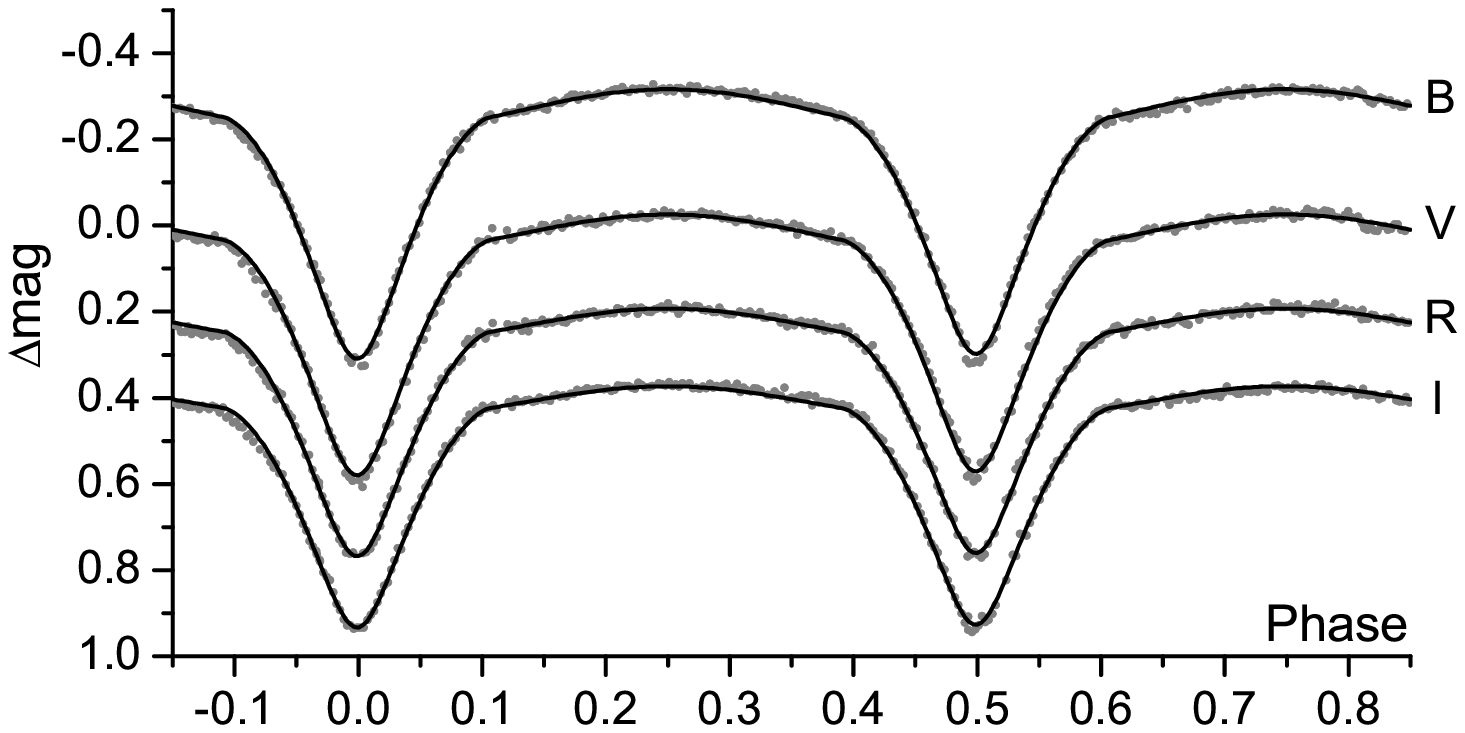}&(a)\\
\includegraphics[width=8cm]{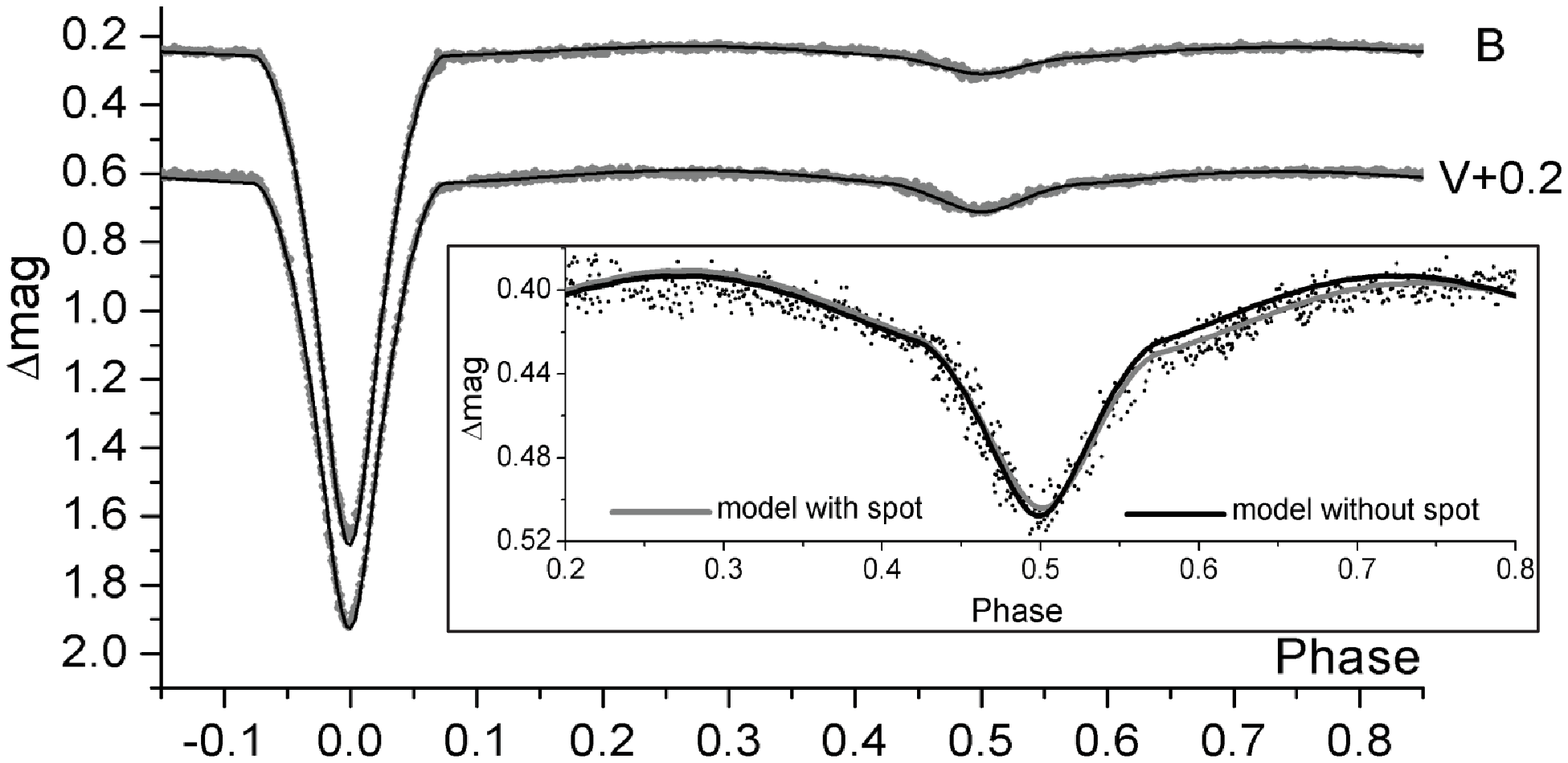}&(b)\\
\includegraphics[width=8cm]{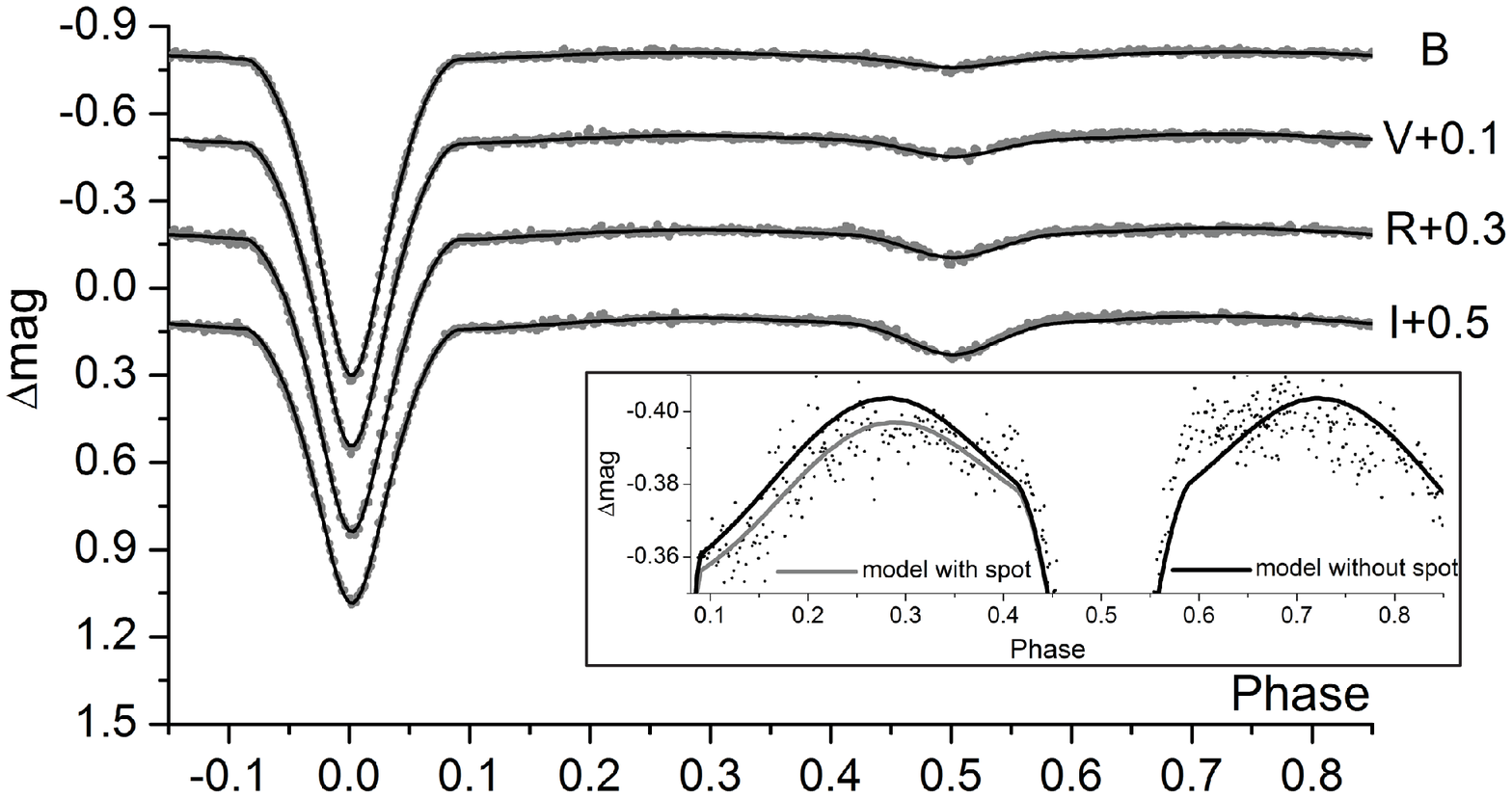}&(c)
\end{tabular}
\caption{Synthetic (solid lines) and observed (points) light curves of (a): AD~And, (b): AL~Cam and (c): V338~Her. For AL~Cam ($V$-filter) and V338~Her ($I$-filter), the comparison models with and without spot assumption are also included.}
\label{fig2}
\end{figure}

We analysed simultaneously complete LCs of each system using all individual observations, with the \emph{PHOEBE} v.0.29d software (Pr\v{s}a \& Zwitter \cite{PZ05}) that follows, more or less, the method of the 2003 version of the Wilson-Devinney (WD) code (Wilson \& Devinney \cite{WD71}; Wilson \cite{WI79,WI90}). In the absence of spectroscopic mass ratios, the `$q$-search' method was applied in modes 2 (detached system), 4 (semi-detached system with the primary component filling its Roche lobe) and 5 (conventional semi-detached binary) to find feasible (`photometric') estimates of the mass ratio. This value of $q$ was then set as an adjustable parameter in the subsequent analysis. The temperature of the primaries $T_1$ were assigned values (9800~K for AD~And, 7600~K for AL~Cam, and 7000~K for V338~Her) according to their spectral class from the tables of Cox (\cite{CO00}) and were kept fixed during the analysis, while the temperatures of the secondaries $T_2$ were adjusted. The Albedos, $A_1$ and $A_2$, and gravity darkening coefficients, $g_1$ and $g_2$, were set to generally adopted values for the given spectral types of the components (Rucinski \cite{RU69}; von Zeipel \cite{VZ24}; Lucy \cite{LU67}). The (linear) limb darkening coefficients, $x_1$ and $x_2$, were taken from van Hamme (\cite{VH93}). The dimensionless potentials $\Omega_{1}$ and $\Omega_{2}$, the fractional luminosity of the primary component $L_{1}$, the system's orbital inclination $i$, and the relative luminosity contribution \emph{$l_{3}$} of a possible third light were set as adjustable parameters. In cases where LC asymmetries were detected, parameters of a matching photospheric spot (latitude, longitude, radius, and temperature factor) were also found. Moreover, the model LCs output from Phoebe were then compared to LCs generated using \emph{CURVEFIT} (cf. Budding \& Demircan \cite{BD07}, Ch.~9), and the results were quite similar for all systems. The observations and their modelling for each system are illustrated in Fig.~\ref{fig2}, and corresponding parameters are listed in Table \ref{tab2}.

\begin{table}
\centering
\caption{The parameters derived from LC fitting.}
\label{tab2}
\begin{tabular}{l ccc}
\hline
\hline
Parameter                 &     AD And      &         AL Cam  &    V338 Her     \\
\hline
Mode                      &    Detached	    &   Semidetached  &	 Semidetached   \\
\hline
$i$ ($^\circ$)            &	    82.6~(3)	&    83.3~(4)	  &	  82.1~(2)      \\						
$q~(m_2/m_1$)	          &	  0.97 (1)      &    0.21~(1)     &   0.27~(1)      \\	
$T_1$ (K)	              &	  9800$^{a}$	&   7600 (170)	  &   7000 (130) 	\\
$T_2$ (K)	              &   9790~(45)	    &   4520~(32)     &   3994~(68)     \\
$BC_1$ (mag)	          &	    $-$0.21	    &      $-$0.03	  &  	 0.00	    \\
$BC_2$ (mag)	          &	    $-$0.19	    &	   $-$0.34    &	   $-$0.68	    \\
$F^{\prime}_{\mathrm{1, V}}$&	 3.970	    &	   3.878	  &	    3.844	    \\
$F^{\prime}_{\mathrm{2, V}}$&    3.968	    &	   3.621      &	   3.524	    \\
$\Omega_1$	              &	   4.31~(4)     &     4.83~(6)    &	    4.05~(2)    \\
$\Omega_2$	              &	   4.14~(3)     &       2.27      &	    2.41	    \\
$A_1^{a}$                 &         1       &        1        &       1         \\
$A_2^{a}$                 &         1       &       0.5       &       0.5       \\
$g_1^{a}$                 &         1       &        1        &       1         \\
$g_2^{a}$                 &         1       &      0.32       &      0.32       \\
$x_{\mathrm{1,B}}$	      &	    0.492	    &	   0.568	  &	     0.595	    \\
$x_{\mathrm{2,B}}$	      &	    0.493	    &	    0.943	  & 	 0.947	    \\
$x_{\mathrm{1,V}}$	      &	    0.418	    &	   0.491	  &	      0.494	    \\
$x_{\mathrm{2,V}}$	      &	    0.419	    &	    0.794	  &	    0.815	    \\
$x_{\mathrm{1,R}}$	      &	    0.353	    &        --	      &	    0.417	    \\
$x_{\mathrm{2,R}}$	      &	    0.354	    &	     --	      &	    0.734	    \\
$x_{\mathrm{1,I}}$	      &	    0.281	    &        --	      &	    0.343	    \\
$x_{\mathrm{2,I}}$	      &	    0.281	    &	     --	      &	    0.601	    \\
$(L_1/L_\mathrm{T})_\mathrm{B}$&  0.475~(3) &     0.930~(4)	  &	  0.931~(3)	    \\
$(L_2/L_\mathrm{T})_\mathrm{B}$&  0.516~(3)	&	  0.070~(4)	  &	   0.030~(2)	\\
$(L_3/L_\mathrm{T})_\mathrm{B}$&0.009~(4)   &	     --	      &	   0.039~(1)    \\		
$(L_1/L_\mathrm{T})_\mathrm{V}$&  0.472~(3) &	   0.886~(4)  &	   0.910~(4)	\\
$(L_2/L_\mathrm{T})_\mathrm{V}$& 0.514~(3)	&	  0.114~(4)	  &	   0.056~(2)    \\
$(L_3/L_\mathrm{T})_\mathrm{V}$& 0.014~(4)  &        --	      &	   0.034~(1) 	\\		
$(L_1/L_\mathrm{T})_\mathrm{R}$& 0.460~(3)  &	    --	      &   0.893~(4)     \\
$(L_2/L_\mathrm{T})_\mathrm{R}$& 0.500~(3)	&	   --	      &	  0.083~(2)	    \\
$(L_3/L_\mathrm{T})_\mathrm{R}$&0.040~(4)   &	     --	      &	   0.024~(2)	\\	
$(L_1/L_\mathrm{T})_\mathrm{I}$&  0.457~(4)	&      --	      &	   0.857~(4)	\\
$(L_2/L_\mathrm{T})_\mathrm{I}$&  0.497~(3)	&	     --	      &	   0.120~(4)    \\
$(L_3/L_\mathrm{T})_\mathrm{I}$&0.045~(5)   &       --        &    0.023~(2) 	\\		
\hline
$L_3$ (\%)	              &	   2.7~(4)	    &	    --	      &	    3.0~(2)     \\							
\hline
                        \multicolumn{4}{c}{\textsl{Spot parameters}}            \\
\hline
Lat~($^\circ$)            &         --    	&	   60~(11)	  &	   89~(30)	    \\
Long~($^\circ$)  	      &	         --	    &	   342~(8)	  &	   105~(5)	    \\
$R$~($^\circ$)     	      &	        --	    &	   18~(2)	  &	    26~(7)	    \\
$T_{\mathrm{spot}}/T_{\mathrm{sur}}$& --    &	   0.8~(1)	  &	    0.9~(1)	    \\
\hline									
$\sum res^{2}$	          &	   0.062        &	   0.072      &	    0.098	    \\
\hline
\multicolumn{4}{l}{$^a$assumed, $L_{\mathrm{T}}= L_1+L_2+L_3,~L_3=4\pi l_3$}
\end{tabular}
\end{table}

Given the brightness level inequality of the LC maxima (`O'Connell effect') of AL~Cam, a cool photospheric spot on the secondary's surface was included in the model providing a fit improvement of $\sim$15\%. In Fig.~\ref{fig2}b, we provide a comparison graph between spotted and unspotted models.

The difference in maximum brightness levels and phases of V338~Her led us to consider the (programmable) options for photospheric inhomogeneities. A cool spot on the secondary can allow the non-uniformity of the LC to be accounted for, and perhaps the consequent inference of magnetic activity of that star could have further implications for the observed cyclic changes of period (see Sect.~7). However, the LC non-uniformity might also be caused by a hot spot on the primary, and such a hot spot would be of suitable brightness to be reasonably associated with the transfer of mass that is also evident in this binary. This point is considered in greater depth in Sect.~9. However, since the temperature of the primary is close to the radiative/convective threshold, both assumptions were tested in the programme. The comparison of the $\Sigma res^2$ values showed that the radiative envelope model provides a more accurate description of the LCs, therefore it is adopted as the final solution.

\section{Absolute elements}

Although no radial velocity curves exist for the systems studied, we can form fair estimates of their absolute parameters. We inferred the mass of each primary from its spectral type, following the correlation of Niarchos \& Manimanis (\cite{NM03}) (see also, Budding \& Demircan \cite{BD07}, Ch.~3), while secondary masses follow from the determined mass ratios. The semi-major axes $a$, which fix the absolute mean radii, can then be derived from Kepler's third law. The parameters are given in Table~\ref{tab3} with formal errors indicated in parentheses alongside adopted values.

\begin{figure}
\centering
\includegraphics[width=8.5cm]{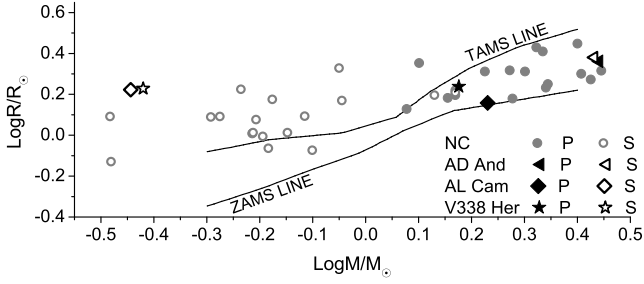}
\caption{The location of the components (P for primary and S for secondary) of the systems in the $M-R$ diagram. The primaries and the secondaries belonging to near contact (NC) systems are indicated. The lines shown are solar-metalicity terminal age main sequence (TAMS) and zero age main sequence (ZAMS), taken from Niarchos \& Manimanis (\cite{NM03}).}
\label{fig3}
\end{figure}

The close binary AD~And shows two very similar A0 type stars both still within the main sequence (MS) band (see Fig.~\ref{fig3}), although with radii $\sim$10\% greater than average. Both AL~Cam and V338~Her are classical Algols. The positions of the systems' components in the $M-R$ diagram (Fig.~\ref{fig3}) are comparable to those of the sample of Niarchos \& Manimanis (\cite{NM03}), which incorporated the data of Russell (\cite{RU48}).

With known absolute radii, it is possible to calculate photometric parallaxes ($\Pi$) using the relation (Budding \& Demircan \cite{BD07}, Ch.~3)
\begin{equation}
\log \Pi = 7.454 - \log R - 0.2 V - 2 F^{\prime}_{\mathrm{V}} \,\,\,   ,
\end{equation}
where the $V$ magnitudes are given in Table~\ref{tab1} and the surface fluxes $F^{\prime}_{\mathrm{V}}$ (cf. Barnes \& Evans \cite{BE76}), following from the known temperatures and corresponding bolometric corrections $BC$, are given in Table~\ref{tab2}. Results are shown in Table~\ref{tab3}, where the agreement between the independently derived distances $D$ of primary and secondary components is a good indication of reliability. The agreement should actually be closer for the primaries, for which magnitudes and colours (and therefore fluxes) are derived with higher precision.

\begin{table}[h]
\caption{The absolute parameters and magnitude estimates of the components of the systems.}
\label{tab3}
\centering
\begin{tabular}{l c c c}
\hline
\hline
Parameter               &  AD And   &     AL Cam    &    V338 Her     \\
\hline
$M_1~(M_{\sun}$)       	&	2.76	&     1.7	    &	   1.5	      \\
$M_2~(M_{\sun}$)       	&2.70 (8) 	&	 0.36 (2)   &	 0.41 (1)	  \\
$R_1~(R_{\sun}$)       	&	2.3 (2)	&    1.4 (1)	&	  1.7 (1)     \\
$R_2~(R_{\sun}$)       	&  2.4 (2) 	&	 1.7 (1) 	&	  1.7 (1)  	  \\
$L_1~(L_{\sun}$)       	&  44 (5) 	&    6 (1)  	&	 6 (1)        \\
$L_2~(L_{\sun}$)       	&  47 (6)  	&	1.0 (2) 	&     0.7 (1)     \\
$a_1~(R_{\sun}$)     	&  3.7 (5) 	&	  1.1 (5)	&	 1.4 (3)      \\
$a_2~(R_{\sun}$)     	&  3.8 (1) 	&	 5.5 (2)    &    5.0 (4)  	  \\
$D_1$ (pc)             	&	1170	&	 319	    &	    286	      \\
$D_2$ (pc)             	&  1200	    &	 302        &	   283	      \\
$M_{\mathrm{1, bol}}$ (mag)&1 (2) 	&	 3 (1) 	    &	  3 (1)       \\
$M_{\mathrm{2, bol}}$ (mag)&1 (2) 	&   5 (1) 	    &	  5 (1)	      \\
$B_1$ (mag)             &	11.81	&	   10.60	&	  10.47	      \\
$B_2$ (mag)             &   12.11	&	  13.40     &     13.86	      \\
$V_1$ (mag)             &	11.68	&	    10.34	&     10.12	      \\
$V_2$ (mag)             &	11.83	&	   12.56	&     13.41	      \\
$R_1$ (mag)             &	11.70	&      --   	&	   9.94	      \\
$R_2$ (mag)             &   11.93	&	   --       &	  12.55       \\
$I_1$ (mag)	            &	11.65	&	   --   	&	  9.82	      \\
$I_2$ (mag)	            &	11.71	&	   --    	&	 12.01        \\
\hline
\end{tabular}
\end{table}

\section{Search for pulsations}

Since two of the systems of the present study, AL~Cam and V338~Her, are candidates containing a pulsating component (Soydugan et al. \cite{SO06}), we performed a short-period pulsation search. To achieve this, the theoretical light curves in all filters were subtracted from the observations and frequencies in the interval 5 to 80~c/d (typical for $\delta$~Sct stars, Breger \cite{BR00}) were checked in the out-of-eclipse data. We used the programme \emph{PERIOD04} v.1.2 (Lenz \& Breger \cite{LB05}), that is based on classical Fourier analysis.

Neither star registered distinct pulsations in the selected range of frequencies with a signal-to-noise ratio higher than 4. However, residuals of V338~Her in the $B$-filter suggest a frequency component (5.9~c/d) near this (4$\sigma$) limit, although its semi-amplitude (1.6~mmag) is very low for the given scatter in the points to be considerable. More accurate measurements are required to help us identify any pulsational behaviour.

\section{O$-$C diagram analysis}

The motion of a distant third star around a close EB causes cyclic changes in the EB orbital period over the timescale of the wide orbit, called the $LI$ght-$T$ime $E$ffect (Woltjer \cite{WO22}; Irwin \cite{IR59}). Computation of the third body's orbital parameters, in this scenario, is a classical inverse problem for five derivable parameters, namely, the period of the wide orbit $P_{3}$, HJD of the periastron passage $T_0$, semi-amplitude $A$ of the LITE, argument of periastron $\omega$, and eccentricity $e_3$. The ephemeris parameters ($JD_0$ and $P$ for the linear form and $C_{2}$ for the quadratic) were calculated together with those of the LITE. The LITE mass function $f(M_3)$ is given as
\begin{equation}
f(M_3) = \frac{1}{P_3^2}\left[\frac{173.145 A}{\sqrt{1 - e_3^2 \cos^2 \omega}} \right]^3 = \frac{(M_3 \sin i)^3}{(M_1+M_2+ M_3)^2}\,
\end{equation}
with the wide orbit's period $P_3$ in yr, and the LITE semi-amplitude $A$ in days. The constant 173.145 is the number of AU traveled at the speed of light in one mean solar day using modern data (cf. Torres et al. \cite{TO10}). The corresponding minimal mass is then $M_{3\mathrm{,min}}=M_3 \sin i_3$ (with $i_3=90^\circ$).
Late-type components of EBs can be expected to have magnetic activity. The observed period changes may therefore be caused by a variation in the magnetic quadrupole moment $\Delta Q$ (Applegate \cite{AP92}). Rovithis-Livaniou et al. (\cite{RO00}) and Lanza \& Rodon\`{o} (\cite{LR02}) proposed that these period changes are given by
\begin{eqnarray}
\Delta P &=& A \sqrt{2 (1 - \cos \{2\pi P/P_{3}\})}\,\,   ,\\
\frac{\Delta P}{P} &=& -9 \frac{\Delta Q}{Ma^2}\,\,   ,
\end{eqnarray}
where $P$ is the close binary period and $M$ is the mass of the magnetically active star. According to Lanza \& Rodon\`{o} (\cite{LR02}), magnetic activity results in a detectable period modulation when $\Delta Q$ is between $\sim10^{50}$~g~cm$^2$ and $\sim10^{51}$~g~cm$^2$. The secondary components of AL~Cam and V338~Her were checked for this.

The ephemerides of the systems were taken from Kreiner et al. (\cite{KR01}) and used to compute, initially, the O$-$C points from all the compiled data. The O$-$C diagram of each system was analysed using an application of the least squares method in a Matlab code (Zasche et al. \cite{ZA09}). Weights were set at $w=1$ for visual, 5 for photographic, and 10 for CCD and photoelectric data. In the cases where more than one minimum timing was available for a given date, their average was used. In Figs~\ref{fig4}-\ref{fig6}, full circles represent times of primary minima and open circles those of the secondary minima, where the larger the symbol, the greater the weight assigned. The corresponding parameters of the solutions are listed in Table~\ref{tab4}.

\subsection{AD And}

\begin{figure}
\centering
\begin{tabular}{c}
\includegraphics[width=8cm]{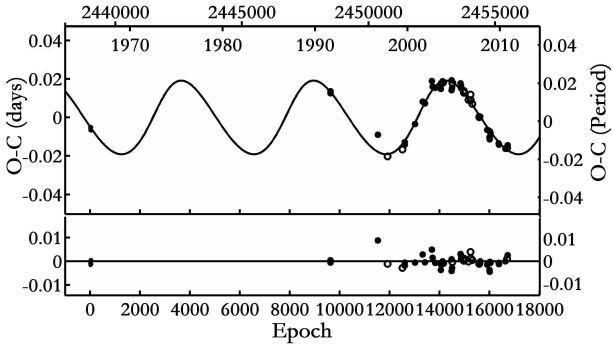}
\end{tabular}
\caption{The O$-$C diagram of AD~And fitted by a LITE curve (upper part) and the residuals after the subtraction of the adopted solution (lower part).}
\label{fig4}
\end{figure}

There are 249 calculated times of minima in the literature since 1910 (93 visual, 103 photographic, 15 photoelectric, and 38 CCD observations). However, most of the O$-$C points before 1990 are of poorer quality and display appreciable scatter. We therefore chose to include 54 minimum timings that are only photoelectric and CCD. The calculated LITE fitting function is shown in Fig.~\ref{fig4}. With our adopted weightings and probable errors, $\chi^2$ was found to decrease from 2000 (linear fit) to 35 after inclusion of the LITE terms, allowing more than 99.9\% confidence in the resulting model improvement (cf. Pearson \& Hartley \cite{PH54}).

\subsection{AL Cam}

\begin{figure}
\centering
\begin{tabular}{cl}
\includegraphics[width=8cm]{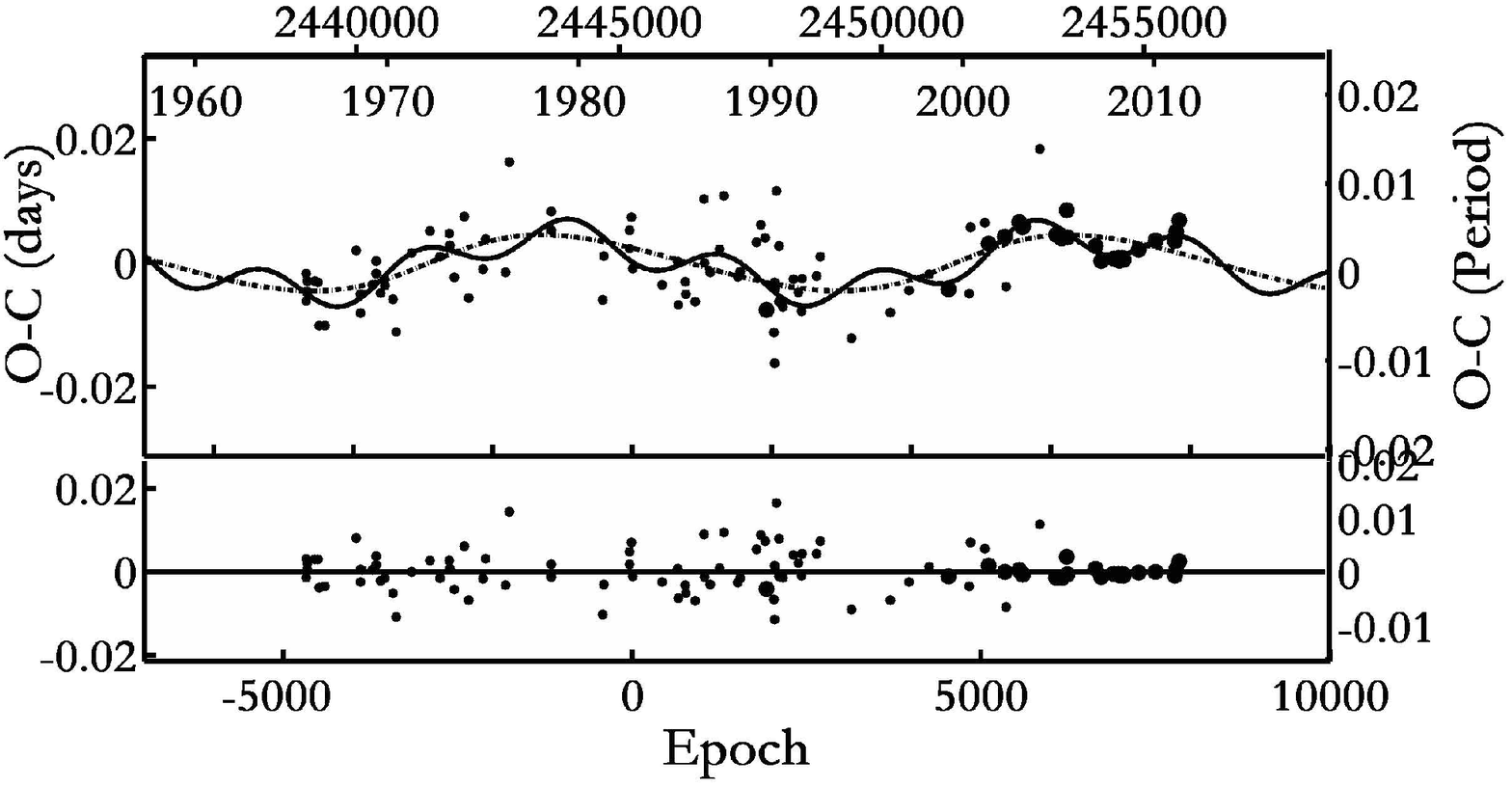}&(a)\\
\includegraphics[width=8cm]{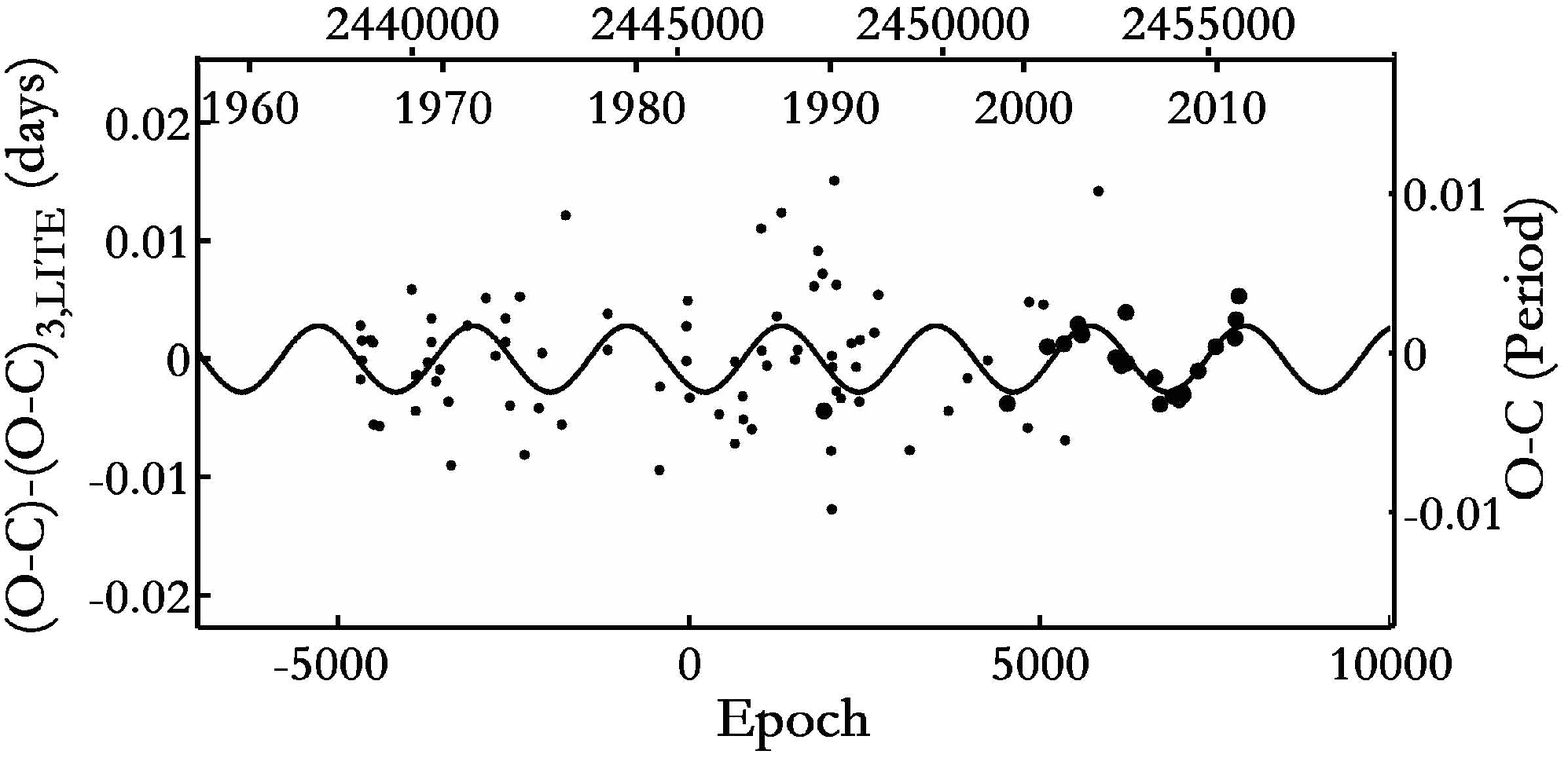}&(b)\\
\end{tabular}
\caption{(a) The O$-$C diagram of AL~Cam fitted by two LITE curves (upper part) and the residuals after the subtraction of the adopted solution (lower part). The solid line indicates the combined fitting terms, while the dashed line corresponds to the first LITE curve. (b) Residuals after removal of the first LITE curve fitted by the second one.}
\label{fig5}
\end{figure}

We used 143 literature times of minima (110 visual, 3 photographic, 5 photoelectric and 25 CCD) to construct the O$-$C diagram of AL~Cam. These data cover an interval from 1965 to the present. Although there are more photographic data before 1950, they were not used owing to their large scatter. A parabolic term (mass transfer indicator) was considered in the fitting function in accordance with the semi-detached model, while in view of apparent low amplitude cyclic variations a LITE function was also included. The parabolic term did not differ significantly from zero, hence was neglected. The fit to the O$-$C points is shown in Fig.~\ref{fig5}a. Owing to the noticeable departures affecting the last few time-of-minimum points, we considered another period component when fitting the LITE residuals. Our corresponding results are shown in Fig.~\ref{fig5}b. However, owing to the small amplitude of this function, which significantly improves only the fits of the most recent data, we cannot be certain that this extra component exists. Future data are required to check this possibility. The parabolic ephemeris reduced $\chi^2$ from 220 (linear fit) to 215, although this dropped to 184 after inclusion of the first LITE term and to 152 after including the second LITE term. This improvement is significant at the 95\% confidence level, while, of course, still leaving a noticeable possibility that the improvement may result purely from chance.

\begin{table*}
\caption{Results of O$-$C diagram analysis.}
\label{tab4}
\centering
\begin{tabular}{l c cc cc}
\hline
\hline
Parameter                           &           AD And     &            \multicolumn{2}{c}{AL Cam}       &          \multicolumn{2}{c}{V338 Her}         \\
\hline
                                    &                               \multicolumn{5}{c}{\textsl{Eclipsing Binary}}                                        \\
\hline
$JD_0$~(HJD-2400000)              	&	  39002.458 (6)	   &       \multicolumn{2}{c}{45252.593 (1)} 	 &	    \multicolumn{2}{c}{33771.369 (9)}  	   	 \\
$P$ (d)                           	&	 0.9861924 (4) 	   &       \multicolumn{2}{c}{1.3283300 (2)} 	 &      \multicolumn{2}{c}{1.305742 (1)}    	 \\
$M_{1}^a + M_{2}^a~(M_{\sun}$)  	&	  2.76 + 2.70  	   &       \multicolumn{2}{c}{1.70 + 0.36}   	 &      \multicolumn{2}{c}{1.50 + 0.41}    	     \\
$C_{2}~(\times10^{-10}$~d/cycle)  	&	      --       	   &           \multicolumn{2}{c}{--}	         &	    \multicolumn{2}{c}{3.404 (1)}    	     \\
$\dot{P}~(\times10^{-7}$~d/yr)    	&	      --       	   &           \multicolumn{2}{c}{--}	         &      \multicolumn{2}{c}{1.904 (1)}     	     \\
\hline
                                    &                                     \multicolumn{5}{c}{\textsl{Mass transfer}}                                     \\
\hline
$\dot{M}~(\times10^{-8}~M_{\sun}$/yr)&	      --           &           \multicolumn{2}{c}{--}	         &      \multicolumn{2}{c}{2.743 (1)}    	     \\
\hline
                                    &                                   \multicolumn{5}{c}{\textsl{LITEs \& additional bodies}}                          \\
\hline
	                                &      $Third~body$    &   $Third~body^b$     &  $Fourth~body^c$   &   $Third~body^b$     &    $Fourth~body^c$     \\	
\hline
$T_0$~(HJD-2400000)	                &	  47012 (175)      &      53677 (7000)    &	      51386 (2000)   &	    33963 (1040)	&	     49111 (658)     \\
$\omega$~($^\circ$)     	        &	     25 (11)       &	     192 (82)     &	        357 (67)   	 &	      49 (94)	    &	      293 (23)     	 \\
$A$~(d)            	                &	     0.019 (1)     &        0.002 (1)  	  &	      0.005 (1)      &	       0.004 (1)	&	     0.014 (1)       \\
$P$~(yr)           	                &	     14.3 (1)      &	     8.0 (2)	  &	         28 (2)	     &	     11.7 (3)	    &	      29.6 (7)       \\
$e$                	                &	    0.17 (5)	   &	    0.04 (1)      &	        0.20 (3)	 &	        0.28 (20)	&	     0.38 (13)       \\
$f(M)~(M_{\sun}$)  	                &	     0.183 (1)     &       0.0010 (1)     &	      0.0010 (3)     &	       0.0021 (1)	&	     0.0155 (2)      \\
$M_{\mathrm{min}}~(M_{\sun}$)	    &	    2.21 (1)	   &	    0.18 (1)      &	        0.15 (1)     &	       0.24 (1)	    &	      0.44 (1)       \\
$a$~(AU)                            &           8.3 (1)    &        4.8 (2)       &         11.1 (1)     &           5.9 (1)    &            11.0 (2)    \\
\hline
                                    &           \multicolumn{5}{c}{\textsl{Quadrupole moment variation in the secondary component}}                      \\
\hline
$\Delta Q~(\times10^{50}$ g cm$^2$)	&	       -- 	       &	      0.68	      &	        0.48	     &	       0.93	        &	         1.40	     \\
\hline
$\sum res^{2}$                    	&	    0.0026   	   &	        \multicolumn{2}{c}{0.0029}	     &			\multicolumn{2}{c}{0.0065}	         \\
\hline
\multicolumn{6}{l}{$^a$The values of $M_{1}$ and $M_{2}$ were taken from Table~\ref{tab3}, based on the $^{c}$first~/~$^{b}$second LITE term}
\end{tabular}
\end{table*}

\begin{figure}
\centering
\begin{tabular}{cl}
\includegraphics[width=8cm]{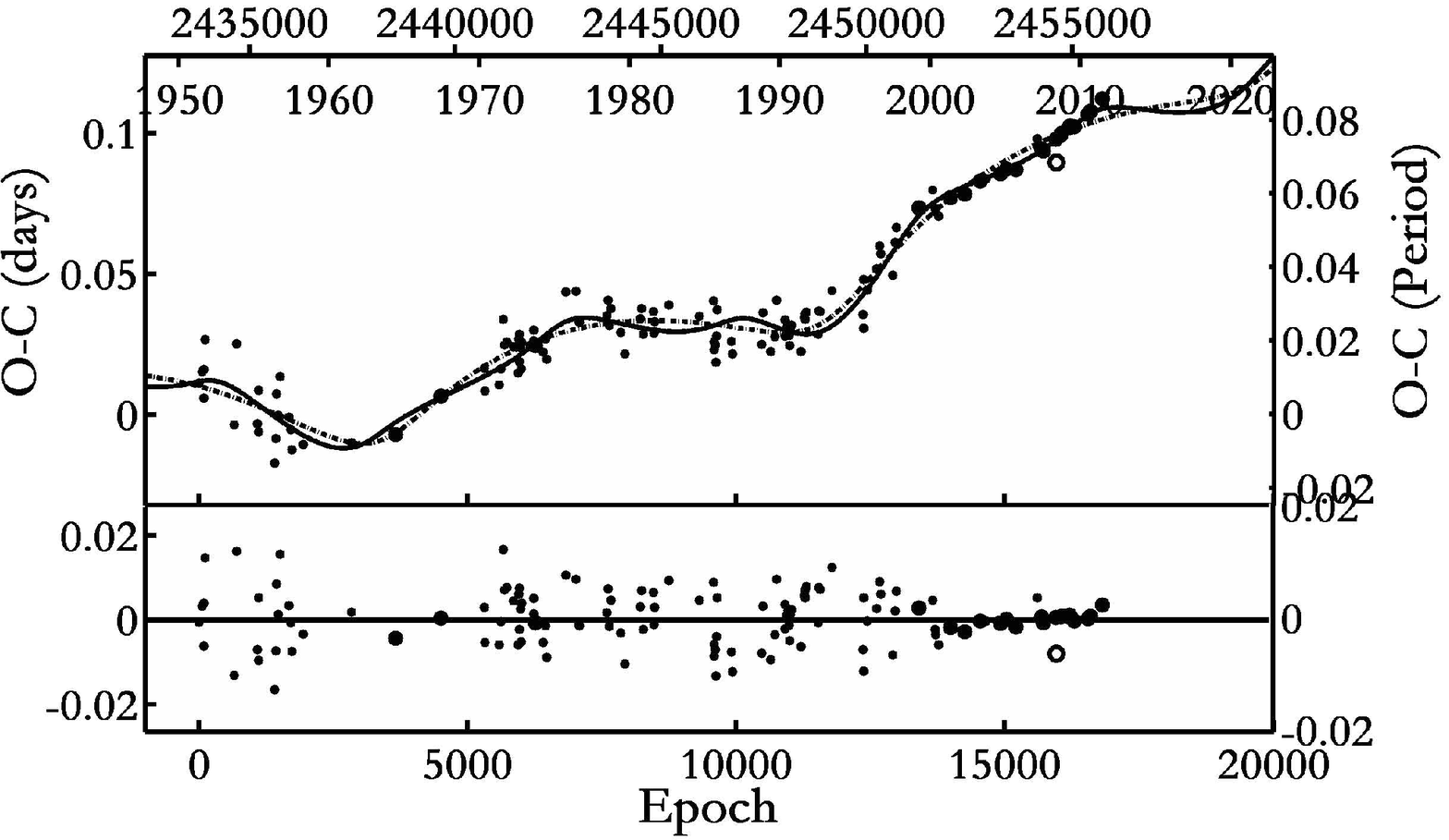}&(a)\\
\includegraphics[width=8cm]{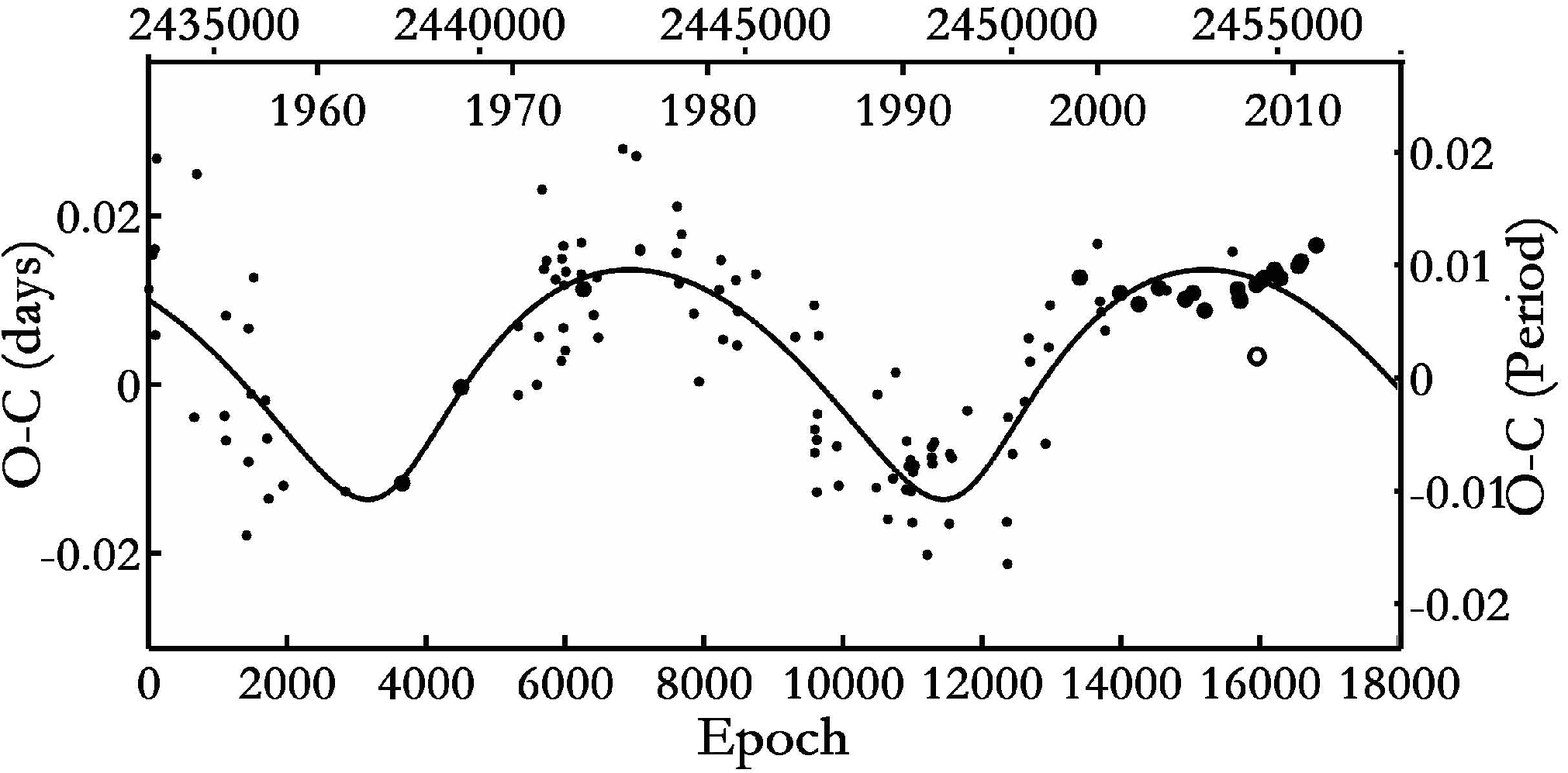}&(b)\\
\includegraphics[width=8cm]{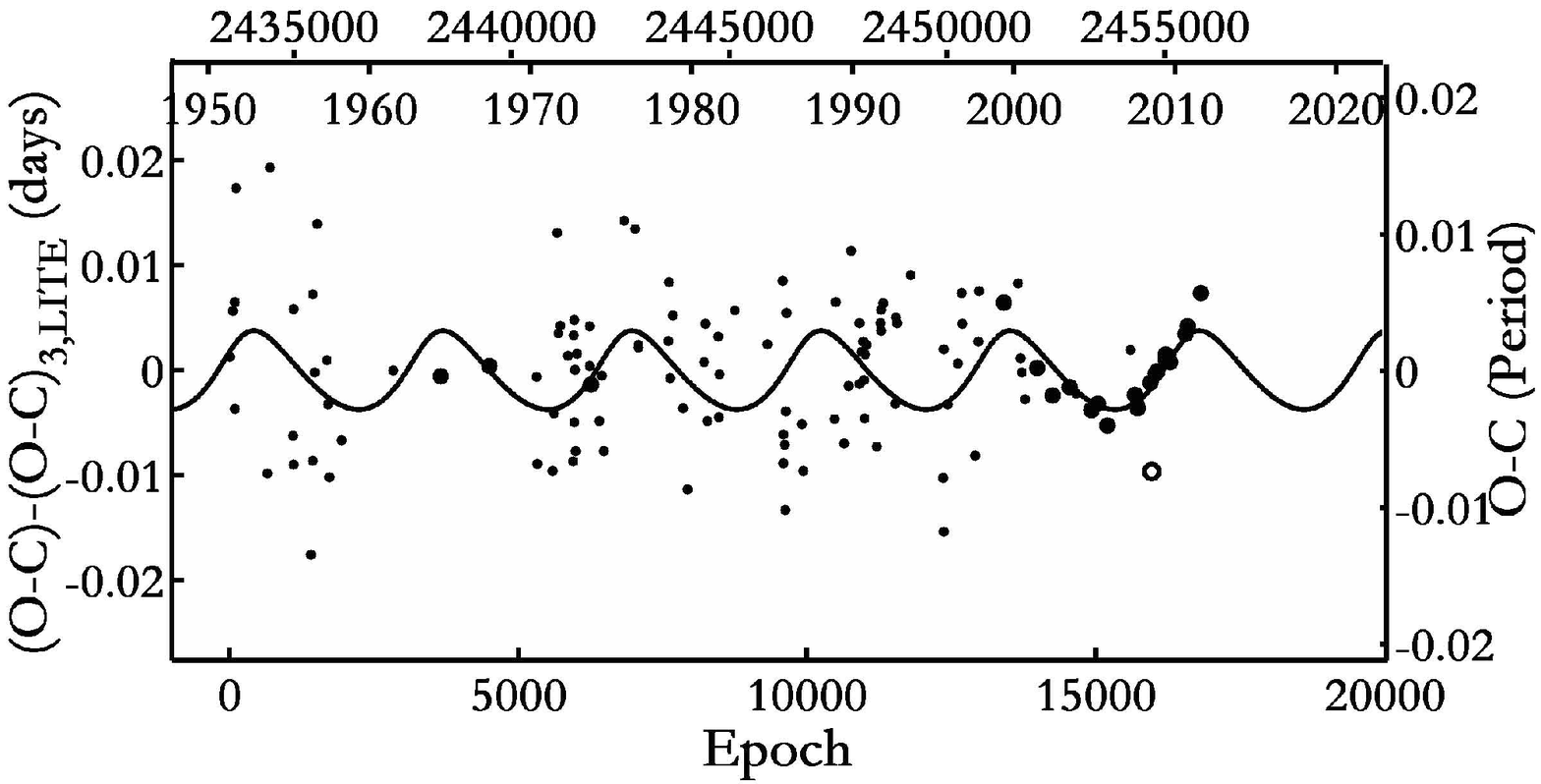}&(c)\\
\end{tabular}
\caption{(a) The O$-$C diagram of V338~Her fitted by two LITE curves and a parabola (upper part) and the total residuals after the subtraction of the complete fitting function (lower part). The solid line indicates the combined fitting terms, while the dashed line corresponds to the parabola and the first LITE function. (b) Residuals after removal of the parabolic term fitted by the first LITE curve. (c) Residuals after removal of both the first LITE curve and the parabola fitted by the second LITE term.}
\label{fig6}
\end{figure}

\subsection{V338 Her}

We collated 161 times of minima from the literature (124 visual, 10 photographic, 6 photoelectric, and 21 CCD). These data come from 1950 up to 2011, with visual observations covering the interval 1950-1995. Both a parabola and a LITE curve were initially included in the modelling. As in the case of AL~Cam, the data of the last decade show additional irregularities, therefore a second LITE curve was added in the final fitting function. The best-fit solution and its residuals are presented in Fig.~\ref{fig6}a, while the LITE solutions alone are shown separately in Figs~\ref{fig6}b-\ref{fig6}c. The $\chi^2$ test showed a decrease from 1530 (linear fit) to 156 for the solution including the parabola and the first LITE term, and to 199 for the final solution. After the inclusion of the second LITE term, the data of the highest quality seem to closely follow the theoretical predictions, with the exception of the last data point, but we are not yet in a position to verify whether another modulating mechanism operates. For this reason, future minima timings will help us to check the present solution.

\section[]{Comparison of LC and O$-$C analyses}

The direction of mass flow, indicated by the sign of the secular period variation, relates to the stellar geometry i.e. the semi-detached condition of a classical Algol. Since the LC fitting also provides fractional luminosities, one can check whether there is a detectable third light contribution $L_{3}$. This is suggested by the O$-$C data for the binaries studied, hence we tested the $L_{3,\mathrm{LC}}$(\%) option in the LC fittings and our findings are listed in Table~\ref{tab2}. The mass-luminosity relation ($L\sim M^{3.5}$ for MS stars) may be compared to the luminosities of the components (Table~\ref{tab3}), and test for the possible luminosity of the third body $L_{3,\mathrm{O-C}}(\%)$ given its minimal mass (Table~\ref{tab4}).

In the case of AD~And, the observed period changes do not show any secular change. This agrees with the derived detached configuration of two MS stars. The LITE model implied a minimal third mass of $\sim2.2~M_{\sun}$, which would yield $L_{3,\mathrm{O-C}}\sim15$\% to the total luminosity of the triple system. This is slight different from the third light ($L_{3,\mathrm{LC}}\sim3$\%) resulting from the LC analysis. This discrepancy can be explained with the assumption that the additional observed light comes from a binary system with two low-mass components instead of one single star. However, the combined findings point to an approximately coplanar multiple star arrangement for AD~And.

Our LC modelling for AL~Cam has implied that it is a semi-detached classical Algol, although no secular increase in orbital period has been found. However, it is known that the observed period behaviour of classical Algols is often more complex than can be interpreted as a conservative mass-transfer process (cf. Budding \& Demircan \cite{BD07}, Ch.~8). In these cases, it has often been assumed that mass transfer could have been interrupted, perhaps by dynamical interactions with a third component, or that other magnetic or systemic effects might be at play. For AL~Cam, however, it is feasible that it is an older, relatively low-mass Algol, whose present rate of mass transfer is too small to be detected in the $\sim$40~yr interval covered by the O$-$C diagram.

The relatively small cyclic period changes experienced by AL~Cam may be associated with one or two additional stars in wide orbits, whose minimal masses would be 0.15~$M_{\sun}$ and 0.18~$M_{\sun}$, respectively, and corresponding minimal light contribution $L_{\mathrm{3+4,O-C}}\sim0.1$\%. This small increase in luminosity would make direct photometric confirmation a difficult task with the given measurement errors. Other methods (e.g. astrometry) may therefore be needed for future clarification of the nature of the additional component(s).

The comparison of the LC and O$-$C results for V338~Her implies that it is a classical Algol based on the LC analysis, a classification that is consistent with the indications of mass transfer inferred from the O$-$C data. For the additional bodies, the O$-$C results provide minimal masses of 0.4~$M_{\sun}$ and 0.2~$M_{\sun}$, and a corresponding relative luminosity $L_{\mathrm{3+4, LC}}\sim3$\%. The minimal $L_{\mathrm{3+4, O-C}}\sim1$\% is thus of approximately the right magnitude to agree with the $L_{3,LC}$ coming from the LC analysis.

For magnetic influences, we derived for AL~Cam $\Delta Q$ values of $\sim5\times10^{49}$~g~cm$^{2}$ and $\sim7\times10^{49}$~g~cm$^{2}$ for the first and second periodic terms, respectively, which are outside the range of the criterion of Lanza \& Rodon\`{o} (\cite{LR02}). The first periodic term of V338~Her resulted in $\Delta Q=1.4\times10^{50}$~g~cm$^{2}$ and the second one in $\sim8\times10^{49}$~g~cm$^{2}$. We may conclude that cyclic period modulations would be difficult to explain with such a low quadrupole moment for AL~Cam. The first periodic term of V338~Her is, however, within the active range and therefore the binary's period might be affected by such a magnetic cycle. Long-term observations of these systems will help us to check for brightness variations related to magnetic cycles.

\section{Discussion and conclusions}

Our combined results indicate that AL~Cam and V338~Her are classical Algols, while AD~And is a detached MS binary. Cyclic effects in the O$-$Cs  of all three systems are indicative of tertiary components. For AD~And, the LITE findings are in marginal agreement with those of Liao \& Qian (\cite{LQ09}). For V338~Her, the classical Algol condition accords with the evidence of mass transfer. The results here are comparable to those of Yang et al. (\cite{YA10}), apart from our argument for a third component. Nevertheless, but we believe that our proposed model is methodologically preferable to that of magnetic influences on the binary period suggested by Yang et al. (\cite{YA10}) in calling for a less \textsl{ad hoc} hypothesis.

AL~Cam might be attended by more than one low mass companion in a wide orbit, but this cannot be confirmed photometrically. Statistical checks of our findings confirmed that they do not provide conclusive evidence of the LITEs, so it would be desirable to acquire future data to determine more precise times of minimum and confirm whether wide orbit scenario is valid.

The dynamical stability of the multiple star configuration for each system was checked by using the stability condition of Harrington (\cite{HA77})
\begin{equation}
\frac{a_{1,2}}{a_3}<\frac{K~\log(3/2)}{\log[1+\frac{m_3}{m_1+m_2}]} \,\,\,   ,
\end{equation}
where $a_{1,2}$ and $a_3$ are the semimajor axes of the binary and the tertiary orbits (see Tables~\ref{tab3}-\ref{tab4}), $m$ the mass of component, and $K$ a constant (=0.28 for co-revolving systems and 0.36 for counter-revolving). This condition is satisfied for both the third and fourth bodies in the cases of AL~Cam and V338~Her. On the other hand, for AD~And we found that, only a counter-revolving configuration satisfies the condition. Hence, if the triple has had a common origin (i.e. the third body was not captured somehow), AD~And should probably be considered as an interesting `non-classical' triple, with a configuration that may well show `interplay', in the sense of Szebehely (\cite{SZ71}). This is a feasible stage of the dynamical evolution of young star systems towards a classical hierarchical arrangement.

AL~Cam and V338~Her were cited as candidates for a pulsating primary. Our the present results have not established this, although for V338~Her there is a weak indication of pulsation in the $B$-filter LC. Future precise photometry, probably using larger telescopes to enable higher time resolution, is needed to follow up our preliminary indications.

Absolute elements of each system were calculated, but these depend on the relevant correlation between primary type and mass for MS stars. The new spectroscopic data agree with previous type classifications (Hill et al. \cite{HI75}) for the two Algol primaries, but the present procedure appears more firmly developed. For AD~And, the spectral type given by Hill et al. (\cite{HI75}) was adopted, and is consistent with our colours. Comparison with Padova model sequences (Marigo et al. \cite{MA08}) confirmed that the primary of V338~Her is below, but close to, a terminal main sequence (TAMS) limit, thus making the adopted correlation self-consistent. On the other hand, the primary of AL~Cam appears close to the ZAMS limit. Both components of AD~And are in the MS band, and, assuming a solar composition for the adopted masses and using the Padova model (Marigo et al. \cite{MA08}), an age of close to 10$^9$~yr can be estimated.
The secondaries of AL~Cam and V338~Her are distinctly overluminous for their mass, which implies that classical Algol systems evolve in the case B scenario. The primary relative radii are also, in both cases, appreciably larger than the conservative-contact limit $r_{1{\mathrm{,lim}}}$ given as $6.08~q^2/(1+q)^4$ (Budding \cite{BU89}), implying that the systems must have evolved non-conservatively, as regards angular momentum and mass transfer. The loser in AL~Cam, for example, must have shed more than three quarters of its mass already, though the period is still only $\sim$1.33~d. In addition, the absence of a detectable secular increase in period implies that the EB is well-advanced in its evolution to an Algol, but perhaps three body dynamical interactions have inhibited the normal expansion of the close pair's orbit (cf. Eggleton et al. \cite{EG07}; Fabrycky \& Tremaine \cite{FT07}). The generally quite similar binary V338~Her, on the other hand, has a period increase $\dot{P}$ of about 1.9~$\times 10^{-7}$~d/yr. It is shown that in the conservative model this would correspond to a mass transfer rate $\dot M~({M_{\sun}}$/yr) (Hilditch \cite{HI01}) of
\begin{equation}
\dot{M} = \frac{\dot{P}}{3P} \frac{M_1~M_2}{(M_1-M_2)} \,\,\,   ,
\end{equation}
where $M_1$, $M_2$ are the masses of the components, and $P$ is the orbital period in days. Even with some degree of non-conservatism in the secondary component's mass-loss process, a rate of mass exchange of around this order is implied by the period increase. This then corresponds to about 2.7~$\times 10^{-8}~M_{\sun}$/yr -- a fairly large value for such a presently low-mass Algol. This mass transfer rate can be related to the LC inhomogeneities of V338~Her (see Sect.~4), since it corresponds to a kinetic energy transfer of about 10$^{31}$ to 10$^{32}$~erg/s in the vicinity of the gainer's photosphere. The LC inhomogeneity would therefore be relatively greater in the $B$ and $U$ wavelength regions, as observed. Even though the photometric consequences of energy transfer can appear at various phases (cf. Richards \cite{RI11}), it would be natural to expect prominent effects corresponding to the post-primary minimum phases, when the initial impact region is exposed. Our failure to find these effects in the present case inhibits a ready preference for the hot spot scenario, which do not therefore adopt. However, a comprehensive model of this mass-transferring Algol, resolving its various physical properties, can be expected to emerge when future data of higher accuracy become available.

The rate of period variation for a given rate of expansion, for a loser of radius $R_2$ (in ${R_{\sun}}$), in the case B model of Algol evolution, can be summarized as (cf. Murad \& Budding \cite{MB84})
\begin{equation}
\frac{\Delta P}{P} = \frac{-9\eta s g(x)}{R_2} \times \frac{P_\mathrm{d}}{365.25} \,\,\,   ,
\end{equation}
where the relative surface density factor $\eta$ and angular momentum coefficient $g(x)$, $x$ being the mass fraction ($q/(1+q)$ in the foregoing notation), can be expected to be similar for the similar configurations of AL~Cam and V338~Her. The rate of loser expansion $s~({R_{\sun}}$/yr), however, depends strongly on the original mass of the loser ($\sim{m_0}^{\mathrm n}$, where $n\approx2.5$). Murad \& Budding (\cite{MB84}) assumed that this should be somewhat greater than half the present total mass of the system, but just how much greater remains an open question for particular systems. There is a clear implication that the original mass of the loser in V338~Her must have been significantly greater than half the present mass of the system (i.e. $\sim1~M_{\sun}$). The clear difference in mass transfer rate between the otherwise presently similar Algols AL~Cam and V338~Her thus points to a hidden difficulty in the unambiguous interpretation of the evolutionary history of classical Algols, arising from lack of information about the proportion of mass lost by the system as a whole. Unless this information can be independently obtained, models of Algol evolution lack independent confirmation when applied to particular cases.

In a general way, the full combination of radial velocity data with photometry is required to securely establish the absolute elements of close binary systems. This has been the well-known `royal road' of eclipse application to stellar astrophysics pioneered by Russell (\cite{RU48}), and is still a pointer for future more detailed studies of stars, similar to those considered in this paper.

\begin{acknowledgements}
This work has been financially supported by the Special Account for Research Grants No 70/4/11112 of the National \& Kapodistrian University of Athens, Hellas. In the present work, the minima database: (http://var.astro.cz/ocgate/), SIMBAD database, operated at CDS, Strasbourg, France, and Astrophysics Data System Bibliographic Services (NASA) have been used. We thank the anonymous reviewer for the valuable comments that improved the quality of the present work. Skinakas Observatory is a collaborative project of the University of Crete, and the Foundation for Research and Technology-Hellas.
\end{acknowledgements}

\listofobjects

\end{document}